\documentclass[%
reprint,
amsmath,amssymb, 
aps,
prl,
]{revtex4-2}
\usepackage{graphicx, color}
\usepackage{braket,appendix} 
\usepackage{dcolumn}
\usepackage{bm}
\usepackage{amsmath}
\usepackage[colorlinks,urlcolor=blue,linkcolor=blue,anchorcolor=blue,citecolor=blue]{hyperref}

\begin{document}
\preprint{APS/123-QED}
\title{Stacking-induced type-II quantum spin Hall insulators with high spin Chern number in unconventional magnetism}
\author{Chao-Yang Tan$^{1,2}$}
\author{Panjun Feng$^{3,4}$}
\author{Ze-Feng Gao$^{1,2}$}
\author{Fengjie Ma$^{3,4}$}
\email{fengjie.ma@bnu.edu.cn}
\author{Peng-Jie Guo$^{1,2}$}
\email{guopengjie@ruc.edu.cn}
\author{Zhong-Yi Lu$^{1,2,5}$}
\email{zlu@ruc.edu.cn}
\affiliation{1. Department of Physics and Beijing Key Laboratory of Opto-electronic Functional Materials $\&$ Micro-nano Devices. Renmin University of China, Beijing 100872, China}
\affiliation{2. Key Laboratory of Quantum State Construction and Manipulation (Ministry of Education), Renmin University of China, Beijing 100872, China}
\affiliation{3.The Center for Advanced Quantum Studies and  School of Physics and Astronomy, Beijing Normal University, Beijing 100875, China}
\affiliation{4.Key Laboratory of Multiscale Spin Physics (Ministry of Education), Beijing Normal University, Beijing 100875, China}
\affiliation{5. Hefei National Laboratory, Hefei 230088, China}
\date{\today}
\begin{abstract}
Generally, stacking two monolayer type-I quantum spin Hall insulators gives rise to a trivial insulator. However, whether or not stacking two type-II quantum spin Hall insulators results in a trivial insulator has not yet been explored. In this letter, based on the calculations of lattice model, we demonstrate that stacking two type-II quantum spin Hall insulators does not yield a trivial insulator, but instead forms a quantum spin Hall insulator with high spin Chern number. In this phase, there are two pairs of topological edge states with opposite chirality and polarization coexisting in the boundary. Our calculations further reveal that the quantized spin Hall conductance of the bilayer is twice that of the monolayer. When U(1) symmetry is present, the high spin Chern number phase remains stable; when U(1) symmetry is broken, it persists over a broad parameter range. Furthermore, based on the first-principles electronic structure calculations, we propose that bilayer Nb$_2$SeTeO is a type-II  quantum spin Hall insulator with high spin Chern number. Finally, extending this strategy to multilayer stacks naturally leads to quantum spin Hall insulator with larger spin Chern number. Our work not only deepens the distinction between type-I and type-II quantum spin Hall insulators, but also offers a route toward realizing highly quantized spin Hall conductance.
\end{abstract}

\maketitle

\begin{figure}[h!]
	\centering
	\includegraphics[width=8.2cm]{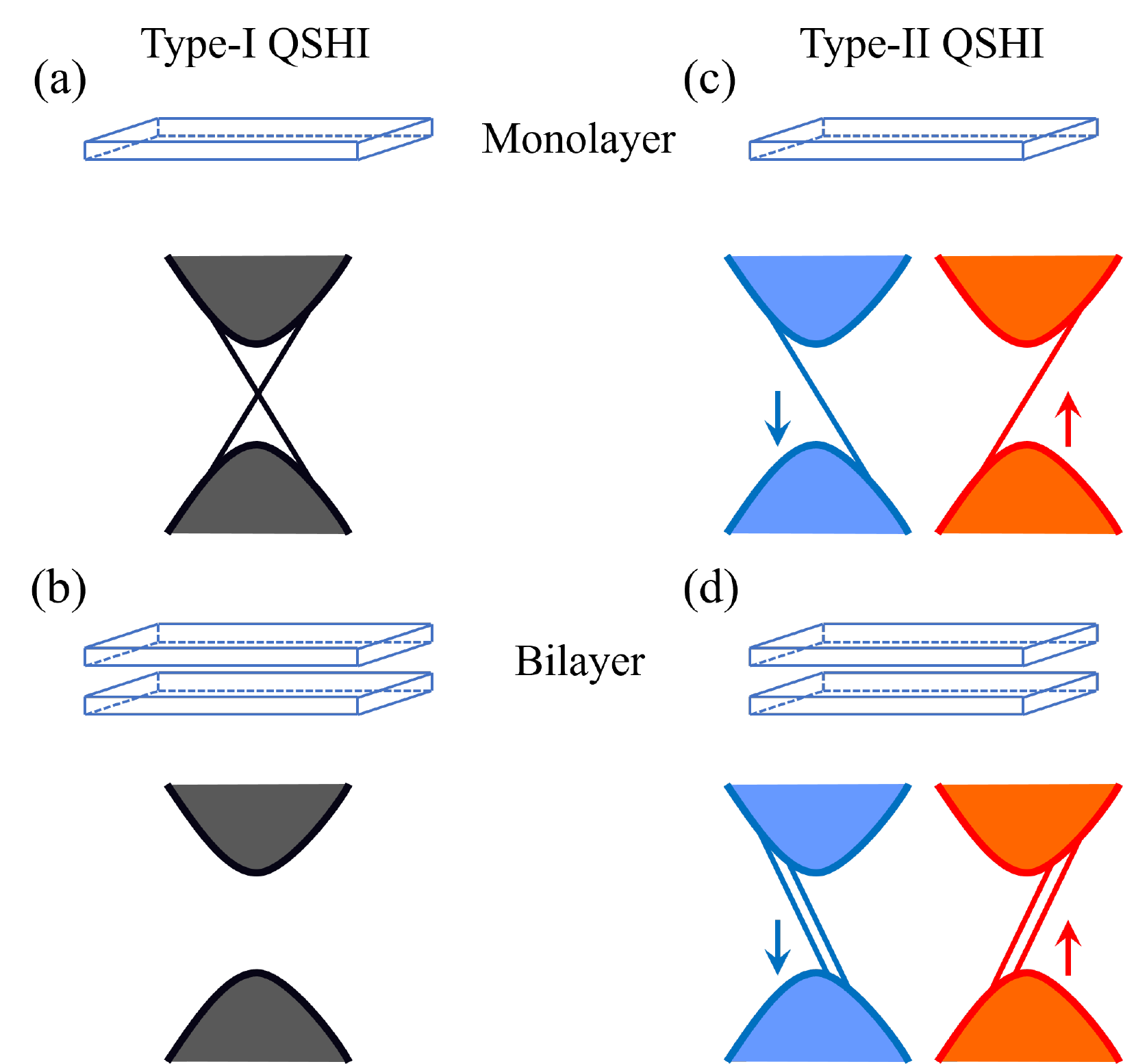}
	\caption{Left panels: the edge states for monolayer (a) and bilayer (b) type-I QSHIs. Right panels: the edge states for monolayer (c) and bilayer (d) type-II QSHIs.}
	\label{fig1}
\end{figure}

\textit{Introduction.} 
The proposal of the conventional (type-I) quantum spin Hall insulator (QSHI) has sparked widespread interest in topological phases and materials \cite{KM-PRL2005-graphene,KM-PRL2005-Z2,QSHE-QWZ,SpinC2006,HgTe-2006,HgTe-2007,TI-RMP2010,2Dto3D-Zhang-PRB,TI-RMP2011,TB-RMP-2016}. A type-I QSHI is known to have a bulk band gap and topological Dirac edge states protected by time-reversal symmetry at the boundary (Fig. \ref{fig1}{(a)}). Its topological invariant is characterized by Z$_2$ index. Stacking two type-I QSHIs typically yields a trivial insulator with Z$_2$ = 0, as illustrated in Fig. \ref{fig1}{(b)}. 

Very recently, type-II QSHI has been theoretically proposed \cite{typeIIQSHI,Fe2Te2O}. Unlike type-I QSHI, the type-II QSHI manifests distinct spin-dependent edge states: its spin-up topological chiral edge state connects the spin-up bulk bands, while the spin-down topological chiral edge state connects the spin-down bulk bands (Fig. \ref{fig1}{(c)}. Because the spin-polarized bands reside in distinct regions of the Brillouin zone (BZ), type-II QSHIs do not require time-reversal symmetry for protection; instead, their topological invariant is characterized by spin Chern number \cite{SpinC2006}. Given that type-II QSHIs necessitate anisotropic spin splitting, they can be realized in altermagnetic and unconventional compensated magnetic materials \cite{typeIIQSHI,Guo-Luttinger,Liu-prl2025,Hou-cpl}.
A natural question then arises: does stacking two type-II QSHIs also yield a trivial insulator? Since type-II QSHIs are classified by spin Chern number, stacking two QSHIs with spin Chern number 1 may result in a type-II QSH insulator with spin Chern number 2. Accordingly, the boundary of such system should host two pairs of edge states with opposite chirality and spin polarization (Fig. \ref{fig1}{(d)}). An important open question is whether or not such a type-II QSHI with high spin Chern number obtained by stacking two type-II QSHIs with spin Chern number 1 can indeed be realized.

In this Letter, we not only demonstrate—via an altermagnetic lattice model—that stacking two type-II QSHIs can yield a type-II QSHI with high spin Chern number, but also predict, from the first-principles electronic structure calculations, that bilayer Nb$_2$SeTeO is such a type-II QSHI with high spin Chern number. The boundary of the bilayer type-II QSHI with high spin Chern number hosts two pairs of edge states with opposite chirality and spin polarization, and its quantized spin Hall conductance is exactly twice that of a monolayer type-II QSHI. Moreover, the bilayer type-II QSHI phase with high spin Chern number remains stable over a wide parameter range, irrespective of whether U(1) symmetry is present.

\begin{figure*}[htbp]
	\centering
	\includegraphics[width=16cm]{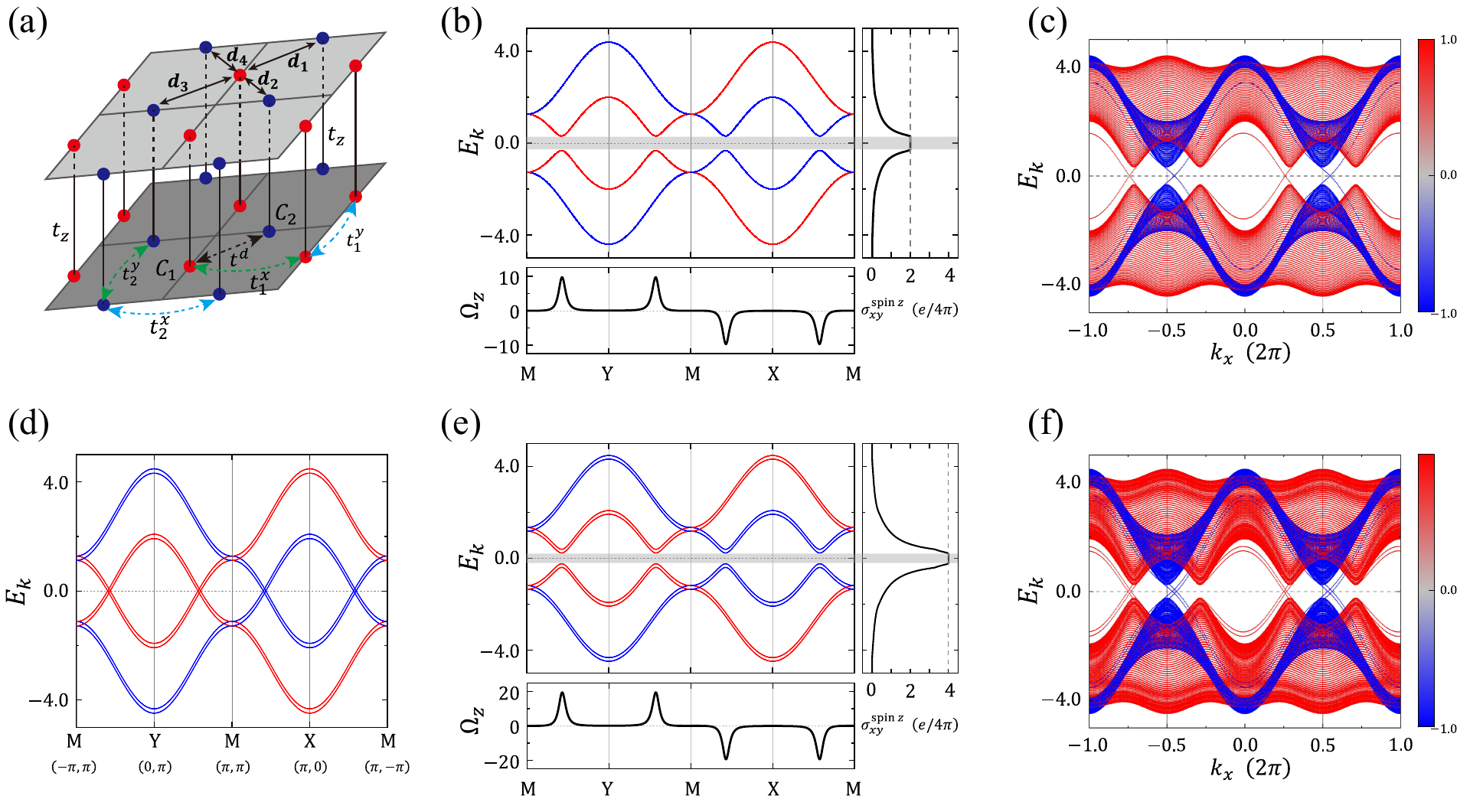}
	\caption{(a) AA-stacked bilayer altermagnetic (AM) lattice model, where sites with different colors represent sublattices carrying opposite magnetic moments. (b) Band structure, Berry curvature and SHC of monolayer AM model with SOC. (c) Corresponding edge states with $s_z$ projection of monolayer AM model, calculated for a 60-layer AM nanoribbon with periodic boundary conditions along $y$-direction and open boundary conditions along $x$-direction. (d) and (e) Band structures for bilayer AM model without ($|\lambda|=0$) and with SOC. The Berry curvature and SHC of bilayer AM model include in (e). (f) Edge states of bilayer AM model under SOC. Parameters: $t^d=1$, $t_1=-t_2=0.8$, $|\lambda|=\lambda_z=0.1$, $|\boldsymbol{m}|=m_z=1.2$ for both monolayer and bilayer AM models, and $t_z=|t_1|/10$ for the bilayer.}
	\label{fig2}
\end{figure*}

\textit{ Lattice Model.} To study the properties of stacking two type-II QSHIs within a bilayer lattice model, we first need to realize type-II QSHI phase on a monolayer lattice model, which we have already constructed in our previous work \cite{QAH-npj2023,BWS,typeIIQSHI}. To make this work easier to understand, we briefly reintroduce the monolayer lattice model. It is a square-lattice model with two sublattices (one layer in Fig. \ref{fig2}{(a)}). The corresponding Hamiltonian reads
\begin{align}
	H_l = &\sum_{d_i,j}\left[ t^{d}C_{1,j}^\dagger C_{2,j+\bm{d}_i}+C_{1,j}^\dagger(i\boldsymbol{\lambda^d}\!\cdot\! \boldsymbol{\sigma})C_{2,j+\bm{d}_i}+\text{h.c.}\right] \notag\\
	&+ \sum_{\alpha,j} \left[ t^{x}_{\alpha} C_{\alpha,j}^\dagger C_{\alpha,j+\mathbf{x}} + t^{y}_{\alpha} C_{\alpha,j}^\dagger C_{\alpha,j+\mathbf{y}} +h.c. \right] \notag\\
	&+ \sum_{\alpha,j}  \left.\bf{m}_\alpha \cdot \boldsymbol{\sigma} \right. C_{\alpha,j}^\dagger C_{\alpha,j} ,
	\label{Eq1}
\end{align}
where $C_{\alpha,j}^\dagger=\bigl( C_{\alpha,j\uparrow}^\dagger, C_{\alpha,j\downarrow}^\dagger \bigr) $ is electron creation operators, $t_d$ and $t_\alpha^{x,y}$ are nearest neighbor hopping along $\mathbf{d}_i$ and next nearest neighbor (NNN) hopping along $\mathbf{x}, \mathbf{y}$, respectively. As illustrated in Fig.~\ref{fig2}(a), these vectors defined as $\bm{d}_{1,4}=\pm\frac{1}{2}(\mathbf{x}+\mathbf{y})$ and $\bm{d}_{2,3}=\pm\frac{1}{2}(\mathbf{x}-\mathbf{y})$, where $\mathbf{x}=a_1 \hat{x}, \mathbf{y}=a_2 \hat{y}$ denote primitive lattice vectors with constants $a_1, a_2$  along unit vectors $\hat{x}, \hat{y}$, respectively.  For simplicity, we set $\boldsymbol{\lambda}^{d_1}=\boldsymbol{\lambda}^{d_3}$, $\boldsymbol{\lambda}^{d_2}=\boldsymbol{\lambda}^{d_4}$, and $\boldsymbol{\lambda}^{d_1}=-\boldsymbol{\lambda}^{d_2}=\boldsymbol{\lambda}$, which represent the parameters of SOC. The $\textbf{m}_\alpha=(-1)^\alpha\textbf{m}$ stands for the static collinear AFM order, $\sigma_0$ and $\boldsymbol{\sigma}$ are identity matrix and Pauli matrix, respectively. 

In this monolayer lattice model, two sites with opposite magnetic moments are respectively located at two space-inversion invariant positions, which endows the model with spin symmetry $\{E || I\}$ while breaking the spin symmetry $\{C^{\bot}_2 || I\}$. Moreover, the NNN hopping satisfies $t_1^{x} \neq t_2^{x}$ and $t_1^{y} \neq t_2^{y}$ but $t_1^{x} = t_2^{y}$ and $t_1^{y} = t_2^{x}$, which breaks spin symmetry $\{C^{\bot}_2 || \tau \}$ while preserving spin symmetry $\{C^{\bot}_2 || C_{4z} \}$ and $\{C^{\bot}_2 || M_{xy} \}$. Consequently, these conditions give rise to $d$-wave altermagnetism in the monolayer lattice model. In addition, this monolayer lattice model has also band inversion by imposing the condition $2|t_1-t_2| > |m|$ with $t_\alpha^x=t_\alpha$ and $t_1=-t_2$. In the absence of SOC, the model can realize a type-I bipolarized Weyl-semimetal phase \cite{BWS}. Upon further including spin-orbit coupling (SOC) , the gap opened in the bipolarized Weyl semimetal drives the model into a type-II quantum spin Hall insulator \cite{typeIIQSHI}.

We first consider $\mathbf{m} \parallel \boldsymbol{\lambda}$ ($\mathbf{m}=m_z \mathbf{z}$) case, the corresponding monolayer lattice model has $U(1)$ symmetry leading to spin $S_z$ being a good quantum number. With SOC, the bipolarized Weyl semimetal transitions into an insulator, as shown in Fig. \ref{fig2}(b). As is well known, a massive Weyl point in two dimensions contributes $\pi$ Berry phase. In this monolayer lattice model, the $C_{2z}$ symmetry ensures that the two Weyl points along the X-M or Y-M boundary contribute the same Berry phase. Specifically, the pair of Weyl points with spin-down polarization on the X-M boundary contribute -2$\pi$ Berry phase, corresponding to Chern number $C_{\downarrow}=-1$, while the pair of Weyl points with spin-up polarization on the Y-M boundary contribute 2$\pi$ Berry phase, giving $C_{\uparrow}=1$. Thus, spin Chern number is $C_{s}=\frac{1}{2}(C_{\uparrow}-C_{\downarrow})=1$. The corresponding spin Hall conductivity (SHC) is quantized as $\sigma_{xy}^{z}=C_{s}e/(2\pi)=e/(2\pi)$, which is consistent with our calculation results (Fig. \ref{fig2}(b)). Moreover, open-boundary-condition calculations further show that there is a pair of topological edge states with opposite chirality and opposite spin polarization along the boundary (Fig. \ref{fig2}(c)). Thus, the type-II QSHI can be realized in the monolayer lattice model. Furthermore, in the case where U(1) symmetry is broken ($\mathbf{m} \nparallel \boldsymbol{\lambda}$) the type-II QSHI phase remains robust over a broad parameter range \cite{typeIIQSHI}.

After realizing the type-II QSHI, we now stack the monolayer lattice model into a bilayer lattice model with interlayer ferromagnetic order and allow weak interlayer hopping $t_z=|t_1|/10$ (Fig. \ref{fig2}(a)). Below we use this bilayer lattice model to investigate the properties of stacked type-II QSHIs. The corresponding Hamiltonian is
\begin{align}
	H = \sum_{l=1}^2 H_l + \sum_{\alpha}\left[ t_z C_{1;\alpha,j}^\dagger C_{2;\alpha,j} + h.c. \right] ,
	\label{Eq2}
\end{align}
where $l=1,2$ and $\alpha=1,2$ denote the layer index and sublattice site index per monolayer, respectively. We define creation operator $C_{l;\alpha,j}^\dagger=\bigl( C_{l;\alpha,j\uparrow}^\dagger, C_{l;\alpha,j\downarrow}^\dagger \bigr) $ for sublattice $\alpha$ of the $l$-th layer. The bilayer lattice model also lacks spin symmetries $\{ C_2^\bot || I \}$ and $\{ C_2^\bot || \tau \}$ and preserves spin symmetries $\{ C_2^\bot || C_{4z} \}$ and $\{ C_2^\bot || M_{xy} \}$. Therefore, the bilayer lattice model is also $d$-wave altermagnetic. 

By performing the Fourier transformation, the Hamiltonian (Eq. \eqref{Eq2}), in absence of SOC, can be rewritten in momentum space as $H=\sum_k \psi^{\dagger} H_k \psi$. The matrix $H_k$ takes the form 
\begin{align}
H_k =\rho_0  \left[  \Gamma^{+}_k \tau_0 + \Gamma_k^{12} \tau_x+\Gamma_k^{-} \tau_z - \tau_z \bf{m} \cdotp \boldsymbol{\sigma} \right]  +  t_z \rho_x \tau_0 
\label{Eq3}
\end{align}
within the basis $\psi^\dagger=(\psi^\dagger_1,\psi^\dagger_2)$, and $\psi^\dagger_l = ( C_{l;1k\uparrow}^\dagger, C_{l;1k\downarrow}^\dagger, C_{l;2k\uparrow}^\dagger, C_{l;2k\downarrow}^\dagger )$ for $l=1, 2$. Here, $\rho_i$ and $\tau_i$ are the Pauli matrices acting on the layer and sublattice, respectively. The auxiliary functions are defined by $\Gamma_k^{12} = 4t^d \cos\frac{k_x}{2} \cos\frac{k_y}{2}$ and $\Gamma_k^{\pm}=(t_1 \pm t_2)\cos k_x + (t_2 \pm t_1)\cos k_y$. The eigenvalues of Eq. \eqref{Eq3} are given by  
\begin{align}
E_{r, s, \pm}(\boldsymbol{k}) = rt_z + \Gamma^{+} \pm \sqrt{(\Gamma_k^{12})^2 + (\Gamma^{-} - s |\boldsymbol{m}|)^2 },
\label{Eq4}
\end{align}
where  $r = \pm 1$ and $s = \pm 1$ denote the layer and spin indices, respectively.

Here,we set the magnetic moment direction along the out-of-plane ($\mathbf{m}=m_z \hat{z}$). As our symmetry analysis shows, the bilayer lattice model remains d-wave altermagnetic and constitutes a bipolarized Weyl semimetal without SOC, as shown in Fig. \ref{fig2} (d). Unlike the monolayer lattice model, there are four pairs of Weyl points at the Fermi level: two pairs with spin-up polarization protected by C$_{2x}$ symmetry, and two pairs with spin-down polarization protected by C$_{2y}$ symmetry. When SOC is considered, C$_{4z}$ T symmetry may drive the bilayer lattice model into a type-II QSHI with high spin Chern number. 

With SOC, Since the bilayer lattice model breaks both C$_{2x}$  and C$_{2y}$ symmetries, all Weyl points open a gap, driving the model from a semimetal to an insulator. As in the monolayer lattice model, we first consider the bilayer lattice model with U(1) symmetry. This bilayer model indeed becomes an insulator (Fig. \ref{fig2}(e)). Calculations of Berry curvature show that the two pairs of massive Weyl points with spin-up polarization contribute identical Berry phases (Fig. \ref{fig2}(e)), so the occupied spin-up states yield a Chern number of 2. Because of the C$_{4z}$T symmetry, the occupied spin-down states contribute a Chern number of –2. Consequently, the bilayer lattice model is a type-II quantum spin Hall insulator with spin Chern number 2; the corresponding spin Hall conductance is $\sigma_{xy}^z = 2e/(2\pi)$, which is confirmed by our calculations (Fig. \ref{fig2}(e)). Moreover, type-II quantum spin Hall insulator with spin Chern number 2 supports two pairs of topological edge states with opposite chirality and polarization on the boundary (Fig. \ref{fig2}(f)).

\begin{figure}[htbp]
	\centering
	\includegraphics[width=8.5cm]{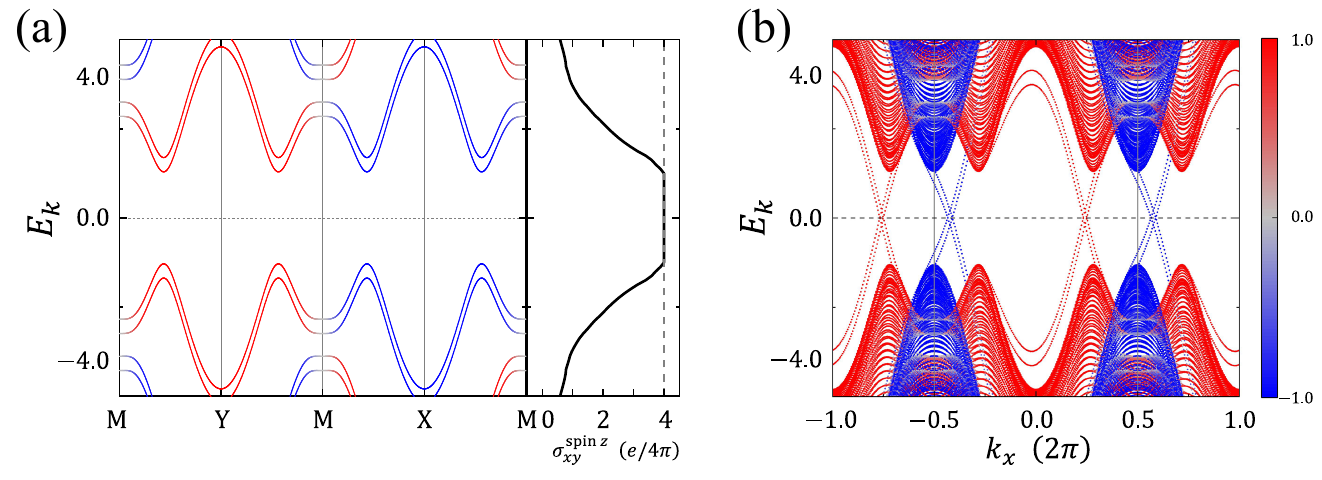}
	\caption{The band structure, SHC and edge state for bilayer AM model at $|\lambda|=0.2$, $|\lambda_x|=0$, $|\lambda_y|=|\lambda|\sin\theta$, $|\lambda_z|=|\lambda|\cos\theta$ with $\theta=\pi/10$. Other parameters are the same with in Fig.\ref{fig2}.}
	\label{fig3}
\end{figure}

In real materials, SOC inevitably breaks U(1) symmetry. We therefore ask whether type-II QSHI with spin Chern number 2 can still be realized once U(1) symmetry is broken. Over a broad parameter range, the bilayer lattice model retains its d-wave altermagnetic character (Fig. \ref{fig3}(a)), and the occupied states of opposite polarization continue to yield opposite Chern numbers of 2 and –2. Moreover, the spin Hall conductance remains quantized to $\sigma_{xy}^z = 2e/(2\pi)$, and open-boundary-condition calculations still show two pairs of topological edge states with opposite chirality and polarization on the boundary (Fig. \ref{fig3}(b)). These results demonstrate that type-II QSHI with high spin Chern number can be realized in materials.

\begin{figure}[htbp]
\centering
\includegraphics[width=8.2cm]{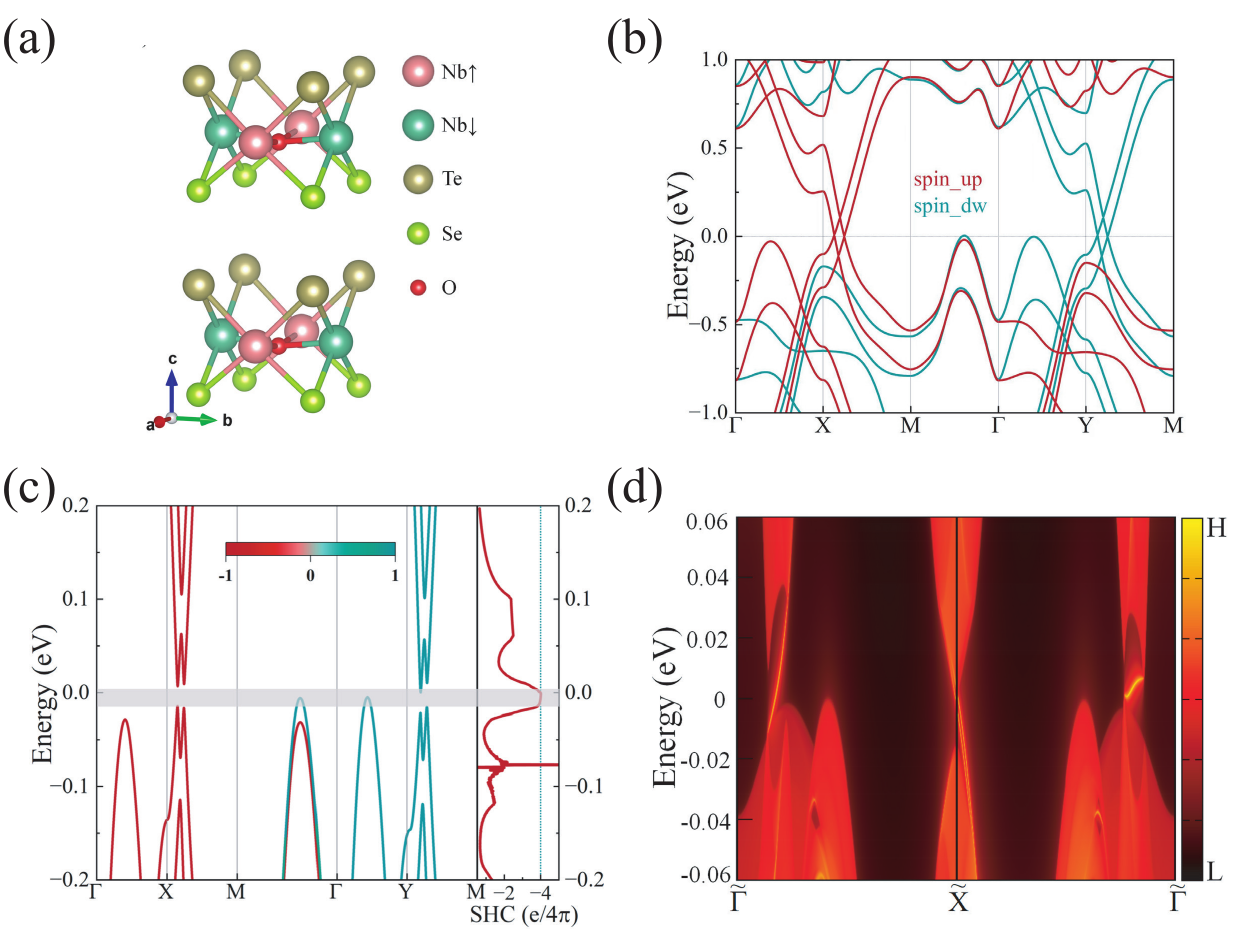}
\caption{(a) Crystal and magnetic structures of the bilayer Nb$_2$TeSeO. (b) The band structure without SOC. (c) The spin resolved band structure and SHC of bilayer Nb$_2$TeSeO. (d) The edge states for bilayer Nb$_2$TeSeO cut along the [100] direction.}
\label{fig4}
\end{figure}

\textit{Candidate materials.} Guided by the model results, realizing type-II QSHI with high spin Chern number by stacking requires that the monolayer is a type-II QSHI. In our earlier work, monolayer Nb$_2$SeTeO was predicted being such a system. The crystal structure of monolayer Nb$_2$SeTeO has the P-4mm space group, and the corresponding point group is C$_{4v}$. As shown in Fig. (Fig. \ref{fig4}(a)) (one layer), it adopts a sandwich structure with Nb–O layer sandwiched between Se and Te sheets, and its magnetic ground state is $d$-wave altermagnetic\cite{typeIIQSHI}. Following the model, bilayer Nb$_2$SeTeO stacked with intralayer antiferromagnetism and interlayer ferromagnetism retains d-wave altermagnetism. Moreover, our calculations indicate that the bilayer Nb$_2$SeTeO is a type-II QSHI with spin Chern number 2. It is however noticed that there is a tiny overlap between the conduction and valence bands. 

To remove this overlap, we apply 0.238 \% tensile strain along the a-axis and  0.238 \% compressive strain along the b-axis. Under this small strain, the symmetry lowers to Pmm2 (point group C$_{2v}$). As a result, the Nb atoms with opposite moments are no longer connected by any symmetry, but the net total moment remains zero, so the strained bilayer Nb$_2$SeTeO transitions from altermagnetism to unconventional compensated magnetism.
We then calculated the electronic structure of the strained bilayer Nb$_2$SeTeO without SOC, shown in Fig. \ref{fig4}(b). The strained bilayer is a bipolarized Weyl semimetal without SOC: spin-up and spin-down Weyl points are protected by M$_y$ and M$_x$ mirror symmetries, respectively. Including SOC, the two pairs of Weyl points with spin-up or spin-down polarization contribute Chern numbers +2 or –2, rendering the strained bilayer Nb$_2$SeTeO to be a type-II QSHI with spin Chern number 2, which is also confirmed by our SHC and edge-state calculations (Fig. \ref{fig4}(c) and (d)). Hence, stacking-induced type-II QSHIs with spin Chern number 2 can be realized in both altermagnetic and unconventional compensated materials. Extending this stacking strategy to multilayers naturally allows type-II QSHIs with larger spin Chern number.

In summary, based on theoretical analysis, we propose that stacking two monolayer type-II QSHIs can give rise to a nontrivial topological phase with high spin Chern number, in stark contrast to type-I systems. It is then demonstrated using lattice model. Moreover, the magnitude of the quantized SHC and the number of topological chiral boundary states of the bilayer system are both twice those of the monolayer. Symmetry analysis combined with the first-principles calculations further predict that bilayer Nb$_2$SeTeO realizes such a QSHI with high spin Chern number, and this holds for both altermagnetic and unconventional compensated magnetism. Extending this strategy to multilayer stacks naturally leads to QSHI with even larger spin Chern number. Our work not only deepens the understanding of type-II QSHIs but also offers a route to realizing quantum spin Hall effects with high spin Chern number.

\begin{acknowledgments}
This work was financially supported by the National Natural Science Foundation of China (Grant No.12434009, No.12204533, No.62476278 and No.12174443), the National Key R$\&$D Program of China (Grant No. 2024YFA1408601),the Fundamental Research Funds for the Central Universities, and the Research Funds of Renmin University of China (Grant No. 24XNKJ15). Computational resources have been provided by the Physical Laboratory of High Performance Computing at Renmin University of China.

C.-Y. T. and P. F. contributed equally to this work.
\end{acknowledgments}

\nocite{*}

\bibliography{Reference}

\end{document}